\documentstyle[12pt,epsf,epsfig]{article}
\def\ga{\mathrel{\raise.3ex\hbox{$>$\kern-.75em\lower1ex\hbox{$\sim$}}}}
\def\la{\mathrel{\raise.3ex\hbox{$<$\kern-.75em\lower1ex\hbox{$\sim$}}}}
\def\be{\begin{equation}}
\def\ee{\end{equation}}
\def\ba{\begin{eqnarray}}
\def\ea{\end{eqnarray}}
\def\ga{\mathrel{\raise.3ex\hbox{$>$\kern-.75em\lower1ex\hbox{$\sim$}}}}
\def\la{\mathrel{\raise.3ex\hbox{$<$\kern-.75em\lower1ex\hbox{$\sim$}}}}

\newcommand{\bi}[1]{\bibitem{#1}}
\newcommand{\fr}[2]{\frac{#1}{#2}}

\newcommand{\omm}{\Omega_m}
\begin{document}

\baselineskip=16pt
\begin{titlepage}
\rightline{UMN--TH--2028/01}
\rightline{TPI--MINN--01/46}
\rightline{McGill 01-21}
\rightline{hep-ph/0110377}
\rightline{October 2001}
\begin{center}

\vspace{0.5cm}

\large {\bf
Evolution of the Fine Structure Constant
Driven by Dark Matter and the Cosmological Constant}
\vspace*{5mm}
\normalsize

{\bf Keith A. Olive$^1$} and {\bf
Maxim Pospelov$^{1,2,3}$}

\smallskip
\medskip

$^1${\it Theoretical Physics Institute, School of Physics and
Astronomy,\\  University of Minnesota, Minneapolis, MN 55455, USA}

$^2${\it Physics Department, McGill University, 3600 University St, \\
Montreal,Quebec H3A 2T8, Canada}

$^3${\it  D\'{e}partement de Physique,
Universit\'{e} du Qu\'{e}bec {\`a}
Montr\'{e}al\\
C.P. 8888, Succ. Centre-Ville, Montr\'{e}al, Qu\'{e}bec,
Canada, H3C 3P8 }

\smallskip
\end{center}
\vskip0.6in

\centerline{\large\bf Abstract}

Bekenstein's model of a scalar field, $\phi$, that affects the
electromagnetic permeability (usually identified with ``changing $\alpha$'')
predicts tiny variations of the effective fine structure constant up to
very high red-shifts, $|\alpha(z=3.5)/\alpha_0-1| < 10^{-10}$,
when the constraints from
E\"otv\"os-Dicke-Braginsky types of experiments are imposed.
We generalize this model by allowing additional couplings of $\phi$ to
both a dark matter candidate and to the cosmological constant.
We show that in a supersymmetric generalization
of Bekenstein's model, the coupling to the LSP, which is assumed to
contribute significantly to the dark matter density, can be up to six
orders of magnitude stronger than the coupling to the baryon energy
density. This allows one to evade the present limits on the
non-universality of the gravitational attraction due to $\phi$-exchange
and at the same time accommodate the effective shift in $\alpha$ at the
level of $\alpha(z=3.5)/\alpha_0-1 \sim 10^{-5}$, reported recently from
observations of quasar absorption spectra.

\vspace*{2mm}

\end{titlepage}

\section{Introduction}

Speculations that fundamental constants may vary in time and/or
space go back to the original idea of Dirac \cite{Dirac}.
Despite the reputable origin, this idea has not received much
attention during the last fifty years
for the two following reasons.
First, there exist various sensitive experimental checks that
coupling constants do not change (See, e.g. \cite{Sister}).
Second, for a long time there has not been any credible
theoretical framework which would predict such changes.

Our theoretical mindset, however, has changed since the advent of the
string theory. One of the most interesting low-energy features of
string theory is the possible presence of a
massless scalar particle, the dilaton, whose vacuum expectation
value defines the size of the effective gauge coupling constants.
A change in the dilaton v.e.v. induces a change in the fine
structure constant as well as the other gauge and Yukawa couplings.
The stabilization of the dilaton v.e.v., which usually renders
the dilaton massive, represents one of the fundamental challenges to be
addressed before string theory can aspire to describe the
observable world. Besides the dilaton, string theory often predicts the
presence of other massless or nearly massless moduli fields, whose
existence may influence particle physics and cosmology and may also
change the effective values of the coupling constants as well.

Independent of the framework of string theory, Bekenstein \cite{Bek}
formulated a
dynamical model of ``changing $\alpha$''. The model consists of a massless
scalar field which has a linear coupling to the $F^2$ term of the $U(1)$
gauge field,
$M_{*}^{-1}\phi F_{\mu\nu}F^{\mu\nu}$, where $M_*$ is an associated mass
scale and thought to be of order the Planck scale.
A change in the background value of $\phi$, can be interpreted as a
change of the effective coupling constant. Bekenstein noticed that $F^2$
has a non-vanishing matrix element over protons and neutrons, of order
$(10^{-3}-10^{-2})m_N$.
This matrix element acts as a source in the $\phi$ equation of motion and
naturally leads to the cosmological evolution of the
$\phi$ field driven by the baryon energy density.
Thus, the change in $\phi$ translates into a change in
$\alpha$ on a characteristic time scale
comparable to the lifetime of the Universe or larger.
However, the presence of a massless scalar field $\phi$ in the theory
leads to the existence of an additional attractive force which does not
respect Einstein's weak universality principle. The extremely accurate
checks of the latter \cite{EDB} lead to a
firm lower limit on $M_*$, $M_*/M_{\rm Pl} >10^3$ that
confines possible changes of $\alpha$ to the range
$\Delta \alpha < 10^{-10}-10^{-9}$
for $0< z <5$ \cite{Bek,Livio}.

This range is five orders of magnitude tighter than the change
$\Delta \alpha/\alpha \simeq 10^{-5}$ indicated in the observations of
quasar absorption spectra at $z=0.5-3.5$ and recently reported
by Webb et al. \cite{Webb01}. Given the potential fundamental importance
of such a result, one should remain cautious until this result is
independently verified. Nevertheless,
leaving aside the issue regarding the reliability of the conclusions
reached by Webb et al. \cite{Webb01}, it is
interesting to explore the possibility of constructing a dynamical model,
including modifications of Bekenstein's model, which could produce a
large change in $\alpha$ in the redshift range $z=0.5-3.5$ and still be
consistent with the constraints on $\Delta \alpha/\alpha$ from
the results of high-precision limits on the violation of equivalence
principle by a fifth force. It is also interesting to study whether the
range $\Delta \alpha/\alpha \simeq 10^{-5}$ could be made consistent with
the limits on $\Delta \alpha/\alpha$ \cite{Oklo}-\cite{LV},
extracted from the analysis
of element abundances from the Oklo phenomenon, a natural nuclear fission
reactor that occurred about 1.8 billion years ago. We note that while
big bang nucleosynthesis provides limits on much longer timescales, these
limits are typically quite weak, $\Delta \alpha/\alpha \sim 10^{-2}$ 
\cite{bbn}.

The gap of five orders of magnitude between the desirable range of
$10^{-5}$ and the bounds of order $10^{-10}$ appear to be insurmountable
for any sensible modification of Bekenstein's theory\footnote{A recent
publication claiming that the $10^{-5}$  change in $\alpha$ is realistic in
this framework \cite{SBM} does not impose the
limits from E\"otv\"os-Dicke-Braginsky experiments.}. In this paper, we
propose a modification of Bekenstein's idea consistent with experimental
constraints, but relies on a large coupling between the non-baryonic dark
matter energy density and the $\phi$ field.

At first, such a coupling may appear strange. Indeed, why should
dark matter interact with the $\phi$ field when it is
known that dark matter particles are not charged \cite{Sacha}
and their electromagnetic form-factors are
also tightly constrained \cite{PtV}? It turns out that in certain classes
of models for dark matter, and in supersymmetric models in particular,
it is natural to expect that $\phi$ would couple more strongly to dark matter
particles than to baryons. It is easy to demonstrate this idea by a simple
supersymmetrization of Bekenstein's interaction. In addition to the
coupling of $\phi$ to the kinetic term,
$F^2$, of the gauge boson, $\phi$ will acquire an additional coupling to
the kinetic term of the gaugino, $M_{*}^{-1}\phi\bar\chi
{\not \! \partial} \chi $. If this gaugino constitutes a significant
fraction of the stable LSP neutralino, as is
often the case, the source of $\phi$ due to the energy density of
dark matter turns out to be dramatically enhanced compared to the
baryonic source,
\be
\fr{{\rm Dark~matter~source}}{{\rm baryonic~source}}
\sim (10^2-10^3) \fr{\Omega_{\rm matter}}{\Omega_{\rm baryon}}
\sim 10^3-10^4.
\label{intro}
\ee
Such an  enhancement factor compensates, although not entirely,
for the tremendous suppression of $\Delta\alpha$ once the
E\"otv\"os-Dicke-Braginsky (EDB) limits on $M_*$ are imposed.
It is then reasonable to study this class
of models in further detail as they are numerically
much more promising than the original Bekenstein framework.

We note that there is another possible ``strategy'' to avoid the EDB
constraint. One can assume the existence of some extremal value $\phi_{\rm
ext}$, in the vicinity of which only $(\phi-\phi_{\rm ext})^2$ couples to
$F^2$. This type of coupling was advocated in Ref. \cite{DP}.
If the cosmological evolution drives $\phi$ close to
$\phi_{\rm ext}$ {\em now} \cite{DP}, i.e. at $z=0$,
the EDB constraints will be relaxed.

We organize this paper as follows. In the next section we generalize
the original Bekenstein model. In section 3,
we solve the field equation for the scalar field $\phi$ and
obtain the predictions for the change of $\alpha$.
In the same section, we impose experimental constraints
and compare the results for $\Delta\alpha$
with the range suggested by Webb et al. \cite{Webb01}.
In section 4, we consider predictions for $\Delta\alpha$
in some specific models and demonstrate one model that passes all
constraints. In section 5, we analyze the class of models with quadratic
couplings to $F^2$. Our conclusions are presented in section 6.

\section{Generalization of Bekenstein's model}

\setcounter{equation}{0}

We start our analysis by formulating a generic action that includes
spin-2 gravity, kinetic and potential terms of a modulus $\phi$,
kinetic terms for the electromagnetic field and baryons
as well as the dark matter action,
\ba
S = \int d^4x \sqrt{-g}\left[- {1\over 2} M_{\rm Pl}^2 R +
{1\over 2} M_*^2 \partial_\mu
\phi\partial^\mu\phi - M_{\rm Pl}^2\Lambda_0 B_\Lambda(\phi)
-{1\over 4}B_F(\phi)F_{\mu\nu}F^{\mu\nu}
\;\;\;\;\;\;\;\;
 \right.\nonumber\\
\left.+\sum_{i=p,n}
\bar N_i (iD\!\!\!\!/
 - m_i B_{Ni}(\phi))N_i+ \fr{1}{2}\bar\chi {\not \! \partial} \chi
-{1\over 2} M_\chi B_{\chi}(\phi)\chi^T\chi\right]+...
\label{action}
\ea
Throughout this paper we assume a $+---$ signature for the metric tensor.
In (\ref{action}), $ M_{\rm Pl} = (8\pi G_N)^{-1/2} =2.4\times 10^{18}
{\rm GeV}$ is the Planck mass and  $M_*$ is its analogue in the
$\phi$ sector. Defined this way, $\phi$ is dimensionless.
$N_i$ stands for neutrons and protons, and
$ D\!\!\!\!/ = \gamma^\mu(\partial_\mu - ie_0 A_\mu)$ for protons
and $ D\!\!\!\!/ = \gamma^\mu\partial_\mu$ for neutrons. Here $e_0$ is the
{\em bare} charge which remains constant throughout the cosmological
evolution (modulo the standard RG evolution of $e_0$ which can be
neglected in our analysis). For definiteness, we assume that the dark
matter is predominantly the non-relativistic Majorana fermion $\chi$.
While it is clear that one can associate $\chi$ with a neutralino,
our approach can be easily generalized to other forms of cold
dark matter. Ellipses stand for the omitted electron and
neutrino terms, as well as for a number of possible interaction terms
(i.e. baryon anomalous magnetic moments, nucleon-nucleon interactions
etc.). All mass and kinetic terms are supplied with $\phi$-dependent
factors denoted $B_i(\phi)$. In this sense, the cosmological constant
term acts as a potential for
$\phi$.

We shall further assume that the change of $\phi$ over
cosmological scales is small, $|\Delta \phi| \equiv |\phi(t=t_0)-\phi(t)|
\ll 1$, where $t_0$ is the present age of the universe. As such,
we can expand all couplings around the current value of
$\phi$, which we choose to be zero, $\phi(t=t_0)=0$,
\ba
 B_\Lambda(\phi) =  1 + \zeta_\Lambda \phi +\fr{1}{2}
\xi_\Lambda \phi^2\nonumber\\\label{expansion}
B_F(\phi) = 1 + \zeta_F \phi + \fr{1}{2}\xi_F \phi^2\\\nonumber
B_{Ni}(\phi) = 1 + \zeta_{i}\phi+ \fr{1}{2}\xi_{i} \phi^2\\\nonumber
B_{\chi}(\phi) = 1 + \zeta_{\chi}\phi+ \fr{1}{2}\xi_{\chi} \phi^2.
\ea
The effective fine structure constant depends on the value of $\phi$. As
such,
$\phi(t)$ and $\Delta \alpha/\alpha$ are directly related,
\ba
   \alpha(\phi) = \fr{e_0^2}{4\pi B_F(\phi)}\nonumber\\
     \label{al-phi}
\fr{\Delta \alpha}{\alpha} = \zeta_F \phi +\fr{1}{2}(\xi_F-2\zeta_F^2)
\phi^2,
\ea
and we have defined $\Delta \alpha/\alpha$ as $(\alpha_0 -
\alpha(t))/\alpha_0$.

The cosmological evolution of $\phi$
follows from the scalar field equation
\be
M_*^2\Box\phi = -  M_{\rm Pl}^2\Lambda_0B^\prime_\Lambda
-B^\prime_F{1\over4}\langle F_{\mu\nu}F^{\mu\nu} \rangle
-\langle B^\prime_n m_n\bar nn + B^\prime_p m_p\bar pp\rangle
- {1\over 2} B_{\chi}^\prime M_\chi
\langle\chi^T\chi\rangle.
\label{master}
\ee
In this formula, primes denote $d/d\phi$, and
the average $\langle...\rangle$ denotes a statistical
average over a current state of the Universe. The term with
$ F_{\mu\nu}F^{\mu\nu}$ can be neglected to a good approximation
as its average is zero for photons, and its contribution
mediated by the baryon density, $\sum_{n,p} n_i
\langle i|  F_{\mu\nu}F^{\mu\nu}|i\rangle$, is already included in the
terms proportional to $B_{n,p}^\prime$. We further note
that for a Dirac fermion $\psi$, the mass term
$m_\psi\bar \psi \psi$ (and the analogous combination for a Majorana
fermion) coincides with the trace of the $\psi$-contribution to
stress-energy tensor, or $\rho_\psi-3p_\psi$. Thus, the only term, that
drives $\phi$ in the radiation domination epoch when $\rho = 3p$
is $\Lambda_0B^\prime_\Lambda$ (see e.g. \cite{nko,tsey}). One can easily
check that the change of $\phi$ induced by this term during radiation
domination will be small compared to the $\Delta\phi$ developed
in the subsequent matter domination epoch. Restricting
eq. (\ref{master}) to matter domination, and assuming a
linearized regime (\ref{expansion}), we derive the following equation
of motion in a Robertson-Walker spacetime with scale factor $a(t)$:
\be
M_*^2(\ddot\phi+3H\dot\phi) = -\rho_m \left(\zeta_{m} + \xi_m\phi\right)
-M_{\rm Pl}^2 \Lambda\left(\zeta_{\Lambda} + \xi_\Lambda\phi\right),
\label{eqphi}
\ee
where  $H= \dot a /a$ and $\zeta_m$ is defined as
\be
\rho_m\zeta_{m} \equiv
\rho_\chi \zeta_\chi + \rho_b(Y_p\zeta_p+Y_n\zeta_n).
\ee
Here, $Y_p$ and $Y_n$ are the abundances of neutrons and protons
in the Universe, including those bound in nuclei. We also assume
that $\rho_m = \rho_\chi +\rho_b$. In a more sophisticated
treatment, one may include the contributions of
electrons, the Coulomb energy stored in nuclei
and other minor effects. As discussed in Refs. \cite{Bek,Livio}, to
good accuracy, $\zeta_m$ remains constant during the
matter dominated epoch.

If the $\phi$-dependent energy density becomes comparable to
$\rho_m$ or $\rho_\Lambda \equiv M_{\rm Pl}^2 \Lambda $, eq.
(\ref{eqphi}) must be solved along with Einstein's equations and energy
conservation as a coupled set of equations. However, the small $\phi$
solutions that we are interested in imply that $\rho_\phi$ is small
and (\ref{eqphi}) can be treated separately, with  $a(t)$ used as an input
function.

\section{Cosmological evolution of the fine structure constant
and the EDB constraint}
\setcounter{equation}{0}

The  cosmological evolution of
$\phi$ can be determined by the $\zeta_i$ terms in eq. (\ref{eqphi})
which becomes
\be
\ddot \phi + 3H \dot \phi = -\fr{1}{M_{*}^2}
\left[\zeta_m \rho_m + \zeta_\Lambda \rho_\Lambda\right ]
=-\fr{\rho_c}{M_*^2}\left[\zeta_m\Omega_m\left({a_0\over a}\right)^3 +
\zeta_\Lambda \Omega_\Lambda\right],
\label{phieq}
\ee
Here $\rho_c = 3H_0^2M_{\rm Pl}^2$ is
the critical density of the Universe
at $t=t_0$ and $\Omega_i=\rho_i/\rho_c$.
The solution to this equation can be easily found \cite{Livio,LV,SBM}.
Throughout this paper we shall assume that the Universe is flat and is
presently dominated by
non-relativistic matter and a cosmological constant, $\Omega_{m}+
\Omega_{\Lambda}=1$. In this case,
the time dependence of the scale factor is given by
\be
a(t)^3=a_0^3 \fr{\Omega_m}{\Omega_\Lambda}
\left[\sinh(\fr{3}{2}\Omega_\Lambda^{1/2}H_0t)\right]^2
\ee
and eq. (\ref{phieq}) can be integrated analytically.
The first integral is given by
\be
\dot\phi = -{3} \omm H_0^2\fr{M_{\rm Pl}^2}{M_*^2}\fr{a_0^3}{a^3}
\left[\zeta_m t + {\zeta_\Lambda \over 4 b} (\sinh(2bt)- 2bt) -t_c\right],
\label{first}
\ee
where $b=\fr{3}{2}\Omega_\Lambda^{1/2}H_0$.
In principle, the constant of integration $t_c$ could be kept arbitrary.
There is, however, only one natural way of fixing it
by imposing initial conditions for $\dot \phi$ deep inside
the radiation domination epoch, i.e. at $t$ close to 0.
As discussed in the previous section, during radiation domination,
the r.h.s of (\ref{phieq}) is effectively zero. This leads to a
$\dot\phi \sim a^{-3}$ scaling behavior and means that any initial value
of $\dot \phi$ will be efficiently damped by the Hubble expansion
over a few Hubble times.
Thus, for the solution in the matter dominated epoch we can safely
take $\dot \phi (t=0) = 0$ or equivalently $t_c =0$.

Integrating (\ref{first}) gives $\phi$ as a function
of time,
\be
\phi(t) = \fr{4}{3}\fr{M_{\rm Pl}^2}{M_*^2}
\left[ (\fr{\zeta_\Lambda}{2}-\zeta_m)(bt_0 \coth(bt_0)- bt\coth(bt))-
\zeta_m \ln\fr{\sinh(bt)}{\sinh(bt_0)}\right].
\label{phit}
\ee

Figure 1 shows three different types of solutions for
$\Delta\alpha/\alpha$
as a function of the red-shift $z$, where $ 1+ z = a_0/a$. In this plot,
we have chosen $\zeta_F=10^{-5}$,
$\Omega_\Lambda = 0.7$ and $\Omega_m = 0.3$.
Comparing the three curves, one can see that
the variation of $\alpha$ at high red-shifts is mostly determined by
$\zeta_m$.
If $\zeta_F$ is negative, one  would need to choose negative
$\zeta_m$ in order to get smaller values
of $\alpha$ in the past. Opposite signs of  $\zeta_F$ and $\zeta_m$ lead to
the larger values of $\alpha$ in the past.

\begin{figure}
\begin{center}
   \epsfig{file=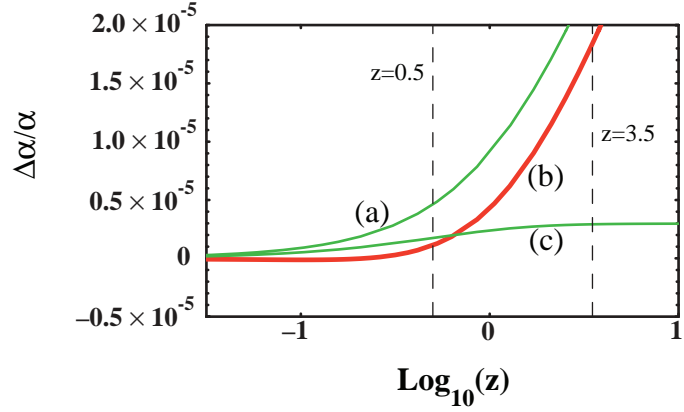,width=9cm,angle=0}
\end{center}
\vspace{0.1in}
 \caption{Three qualitatively different types of solutions for
$\Delta\alpha(z)/\alpha_0$
that give smaller values of $\alpha$ in the past for positive $\zeta_F$. They
correspond to the choice of $\zeta_F = 10^{-5}$ and
(a) $\zeta_m=1$, $\zeta_\Lambda=0$ (b) $\zeta_m=1$, $\zeta_\Lambda=-2$
and (c)  $\zeta_m=0$, $\zeta_\Lambda=1$. The interval of $z$, considered by
Webb et al., $0.5\le z\le 3.5$ is shown by two vertical dashed lines.}
\end{figure}

Given the large parameter space, $(M_*,~\zeta_F,
~\zeta_m,~\zeta_\Lambda)$, one could expect that it is easy to get
$\Delta\alpha(z=0.5-3.5)/\alpha
\sim 10^{-5}$ as suggested by the analysis of the quasar absorption spectra by
Webb et al. \cite{Webb01}.
On the other hand, it is clear that the EDB
constraints should severely restrict the
parameter space of our model.
The differential
acceleration of two elements with different
$A_{1,2}$ and $Z_{1,2}$ towards a common attractor
can be expressed in terms of $\zeta_i$ and $\omega = {M_*^2}/2 {M_{\rm
Pl}^2}$ (See, e.g. \cite{Dicke,Bek}),
\ba
\fr{\Delta g}{\bar g}=2\fr{g(A_1,Z_1)-g(A_2,Z_2)}{g(A_1,Z_1)+g(A_2,Z_2) }
= \fr{1}{\omega}\left(7 \times 10^{-4}\zeta_F
\fr{\overline{Z^{2}/A^{1/3}}}{\bar A}
+\fr{\bar A - \bar Z}{\bar A} \zeta_n + \fr{\bar Z}{\bar A} \zeta_p \right)
\times\nonumber\\
\left[(\zeta_n-\zeta_p)\left(\fr{Z_1}{A_1}-\fr{Z_2}{A_2}\right)+
7 \times 10^{-4}\zeta_F\left(\fr{Z_2^2}{A_2^{4/3}}
-\fr{Z_1^2}{A_2^{4/3}}\right)\right].
\label{deltag}
\ea
where $\bar Z$ and $\bar A$ represent average $Z$ and $A$ of the common
attractor, $\bar Z = \sum n_iM_iZ_i/ \sum n_iM_i$. The terms proportional
to $\zeta_F$ correspond to the electromagnetic contribution to the
total  energy of nuclei.
The best constraints on long-range forces
 are extracted from $\Delta g/\bar g$ measured in experiments
that compare the acceleration of light and heavy elements.
The differential acceleration of platinum and aluminium
is $ \le 2 \times 10^{-12}$ at the 2$\sigma$ level
(last reference in \cite{EDB} as quoted in \cite{Bek}), and the
differential acceleration of the Moon (silica-dominated) and the Earth
(iron-dominated) towards the Sun  is $\le 0.92 \times 10^{-12}$
\cite{LunarR}. Choosing the appropriate values of $Z$ and $A$
and retaining only the  hydrogen contribution to the mass of the Sun, we
get
\ba
\fr{1}{\omega}\left|
\zeta_p(\zeta_n-\zeta_p+2.9\times10^{-2}\zeta_F)\right|<
2.5\times10^{-11} \qquad {\rm Al/Pt ~system}\nonumber\\
\fr{1}{\omega}\left|\zeta_p(\zeta_n-\zeta_p+
1.8 \times10^{-2}\zeta_F)\right|<
2.5\times10^{-11} \qquad {\rm Si/Fe ~system }
\label{EDBl}
\ea
These limits were also considered in a recent paper \cite{DZ}.
$\zeta_n-\zeta_p$ and $\zeta_F$ enter in eqs. (\ref{EDBl}) in different
linear combinations. Thus, it is possible to extract {\em separate} limits
on $\omega^{-1} \zeta_p\zeta_F$ and $\omega^{-1} \zeta_p(\zeta_n-\zeta_p)$.
Models that have non-zero $\zeta_F$ also have non-vanishing $\zeta_{p,n}$
unless some intricate conspiracy of quark, gluon and photon contributions
occur. Barring such possible cancellations, one obtains
$|\zeta_{n,p}| \ga |\zeta_n-\zeta_p|\ga 10^{-3}|\zeta_F|$.
Using these relations, we can
combine the preferred range of Ref. \cite{Webb01} with the constraints,
imposed by eqs. (\ref{EDBl}).

\begin{figure}
\begin{center}
   \epsfig{file=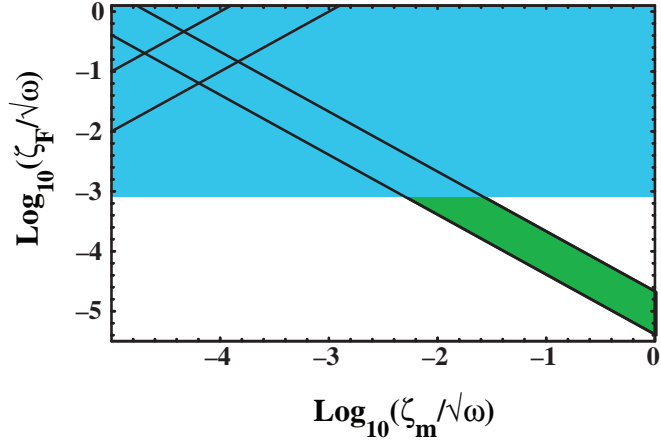,width=9cm,angle=0}
\end{center}
\vspace{0.1in}
 \caption{The $(\zeta_m/\sqrt{\omega},~\zeta_F/\sqrt{\omega})$ parameter
space. The dark-shaded (green) region is consistent with both the EDB
constraints and with a possible relative change of $\alpha$ at the
$10^{-5}$ level, as suggested by Webb et al. \cite{Webb01}. The light
shaded (blue) region is excluded by EDB constraints.
$\zeta_\Lambda$ is set to zero in this plot.}
\label{plane}
\end{figure}

The region excluded by the EDB constraints in
the $(\zeta_m/\sqrt{\omega},~\zeta_F/\sqrt{\omega})$ parameter space
is shown by the light shaded (blue) region in Figure \ref{plane}.
Here we have set $\zeta_\Lambda$ = 0.
The long negative-sloped
band that connects the upper-left and lower-right hand corners is the
range that reproduces $\Delta\alpha/\alpha=10^{-5}$ in the
interval $0.5\le z\le 3.5$. In the original Bekenstein model,
$\zeta_m = (10^{-4}$ to $10^{-3}) \zeta_F$ and corresponds to
the positive sloped band close to  the upper-left corner
\footnote{$\zeta_m = 10^{-3}\zeta_F$ would require rather
``generous'' assumptions
concerning nucleon matrix elements and/or $\Omega_b$.}. As one can see,
the diamond-shaped intersection is deep inside the range {\em excluded} by
the EDB experiments. Of course, this is in agreement with conclusions of
\cite{Bek,Livio}.  Finally, the dark-shaded (green) area represents
the choice of parameters that can reproduce $\Delta\alpha/\alpha=10^{-5}$
\cite{Webb01} and still be in agreement with the EDB constraints. For
this region,
$\zeta_m/\sqrt{\omega}\ga 3\times 10^{-3}$ and
$\zeta_F/\sqrt{\omega}< 10^{-3}$, which points towards models in which
$\phi$ couples to dark matter and the couplings to baryons and $\zeta_F$
are suppressed.

In addition, we must check whether or not these choices of parameters which
satisfy the EDB constraints are also in agreement with limits on
$\Delta \alpha/\alpha$, derived from isotope abundances in the Oklo
natural reactor. Typically, these limits are strong, $|\Delta
\alpha/\alpha| < 1.2\times 10^{-7}$ \cite{DD} and go back to $z\simeq
0.14$ \footnote{The redshift, $z= 0.014$, corresponds to the choice
$\Omega_\Lambda = 0.7, \Omega_m = 0.3, h_0 = 0.65$, and we have assumed
that the Oklo event took place 1.8 Gyr ago.}. This seems to be
dramatically smaller than the range suggested by
\cite{Webb01}. Moreover, there is no way of suppressing
$\Delta\alpha(z<0.14)/\Delta\alpha(0.5<z<3.5)$ below the $10^{-2}$ level
using our freedom in $\zeta_F$ or $\omega$, as these parameters cancel in
the ratio.

There is, however, an extra free
parameter which may be used in an attempt to reconcile a change of
$10^{-5}$ at $0.5\le z \le 3.5$ and the Oklo limit. The behavior of
curve (b) in Figure 1 suggests that $\zeta_\Lambda$ can be used
to make $\Delta\alpha$ almost flat at $z<0.2$.
\begin{figure}
\begin{center}
   \epsfig{file=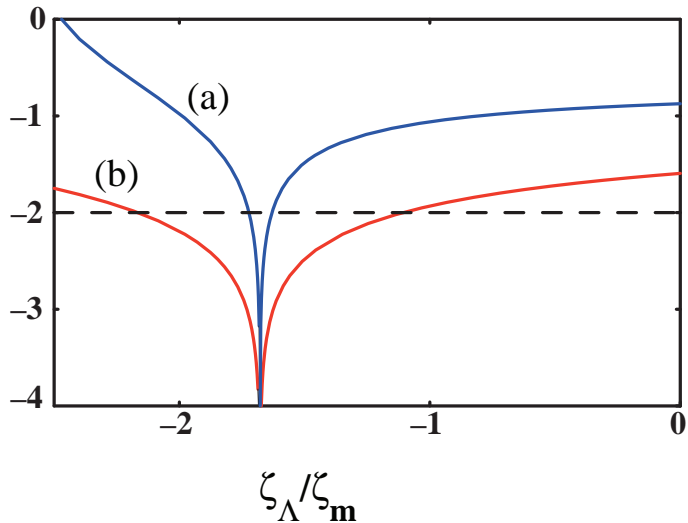,width=9cm,angle=0}
 \end{center}
\vspace{0.1in}
 \caption{ A plot of $\log_{10}(|\Delta\alpha(z =
0.07)/\Delta\alpha(z=0.5)|)$ (a) and
 $\log_{10}(|\Delta\alpha(z = 0.07)/\Delta\alpha(z=3.5)|)$ (b) as a
function of $\zeta_\Lambda/\zeta_m$ for the choice of
$\Omega_\Lambda=0.7$ and
$\Omega_m=0.3$. The portion of the curves below the horizontal
dashed line are consistent with the Oklo limits \cite{DD} and the Webb et
al. suggested change in
$\alpha$. }
\end{figure}
In order to determine the requirements on $\zeta_\Lambda$, we
quantify the comparison between ``Oklo change'' and ``quasar change'' as
follows. In the case of the Oklo constraints, in principle,
one needs to average $\alpha(t)$ over the interval
$0<t_0 - t\la 2\times 10^9$ yr. Since the exact timing of Oklo event is
known only approximately, we choose to quantify it by simply
taking $\alpha$ at the half of the Oklo redshift,
$\Delta \alpha(z=0.07)/\alpha$. This value must be approximately
two orders of magnitude smaller than $\Delta \alpha/\alpha$, suggested by
Webb et al. Thus, we consider the ratio,
$|\Delta \alpha(z=0.07)/|\Delta \alpha(z=0.5)|$ and
$|\Delta \alpha(z=0.07)/|\Delta \alpha(z=3.5)|$ as a function of
$\zeta_\Lambda/\zeta_m$. The logarithms of these ratios  are plotted
in Figure 3.
As one can see, these ratios are two funnel-like curves and it is
possible to choose $\zeta_\Lambda/\zeta_m$ in such a way that
$\Delta\alpha_{\rm Oklo}/
\Delta\alpha_{\rm quasar} < 10^{-2}$. For $z=3.5$ this can be done rather
easily in the range $-2.2<\zeta_\Lambda/\zeta_m < -1.2$.
 For $z=0.5$ one has to choose this ratio rather carefully,
$\zeta_\Lambda/\zeta_m \simeq -1.7 $, and requires
a 5\%  -10\% fine-tuning.
We also note that this specific
value of  $\zeta_\Lambda/\zeta_m $ is very sensitive to
the choice of $\Omega_\Lambda$ and $\Omega_m$ and varies significantly when
$\Omega_\Lambda$ and $\Omega_m$ are varied within their current error bars.
The use of more restrictive bounds from Oklo by \cite{Fujii}
would only worsen the fine tuning.

The above exercise allows us to conclude that in
principle a generalized Bekenstein-like model can yield
$\Delta\alpha/\alpha\sim 10^{-5}$ at $0.5\le z\le 3.5$ and still be in
agreement with the EDB and Oklo constraints. The limits from
non-universality of a fifth force could be evaded in models with
large couplings to dark matter and small couplings to baryons and
$F_{\mu\nu}F^{\mu\nu}$. Also, the Oklo bounds could be avoided or
softened if the dark matter provides a negative push to $\phi$ at later
epoch. Although such a suppression of $\Delta\alpha_{\rm Oklo}$ may
happen, it would appear to be highly accidental.

\section{Model realizations}
\setcounter{equation}{0}

 It is important to note that
neither the original Bekenstein model \cite{Bek} nor its
modifications discussed here are fully defined
at the quantum level. Indeed, the $B_F(\phi)F_{\mu\nu}F_{\mu\nu}$ term
contains not only the bare QED Lagrangian term but also higher dimensional
operators such as $\phi^nF_{\mu\nu}F_{\mu\nu}$. It is clear then that
at the loop level this will create all other possible interactions such as
$\phi^n m_e\bar ee $, $\phi^n m_q\bar qq $, etc., generally, with
divergent coefficients which cannot be fixed from first principles.
While these terms are not expected to drastically change the model
if one makes some plausible assumptions about the cutoff in the
theory, there is, however, the
set of operators contained in $B_\Lambda(\phi)$ which are very sensitive
to the cutoff and are very important as they can give rise
to the mass of the $\phi$ field, an effective cosmological constant, etc.
Unfortunately, the present status of the underlying theory does not allow
for a meaningful calculation of $B_\Lambda(\phi)$. This problem
is, of course, tightly related to the cosmological constant problem
\cite{W}, and/or to the smallness of the mass term for the quintessence
field. As we have nothing to add to these issues, we must
assume that $\zeta_\Lambda$ and $\xi_\Lambda$ are basically
incalculable input parameters and fix $\Lambda$  to its value implied
by the observation of high $z$ supernovae, the anisotropy of the cosmic
microwave background and large scale structure formation.
In what follows, we compile a list of models which predict certain
values for the $\zeta_i$ couplings and/or $\omega$ and confront them with
the phenomenological constraints, discussed in the previous section.

{\em 1 The Bekenstein model}\\
In this model, one initially introduces the coupling of $\phi$ to
$F^2$, $B_F(\phi) = \exp(-2\phi)$  or $\zeta_F=-2$. $B_\Lambda$ can be
set to a constant so that $\zeta_\Lambda = 0$.  In the original model,
$\omega = 1$, however, we will keep it arbitrary for now. The change of
$\phi$ is driven by the electromagnetic fraction of the baryon energy
density. The coupling of $\phi$ to nucleons is given by the same
matrix elements,
$\zeta_{N}=m_N^{-1}\langle N|\fr{\zeta_F}{4}F_{\mu\nu}F^{\mu\nu}|N\rangle
\simeq -m_N^{-1}\langle N|\fr{\zeta_F}{2}(E^2-B^2)|N\rangle$, that determine
the contribution of a ``photon cloud'' to the nucleon mass. Both the Naive
quark model and dispersion approaches give consistent estimates of
these matrix elements \cite{GL}.
Using the results of \cite{GL}, presumably
valid to 50\% accuracy, we find that
$\zeta_p\simeq - 0.0007 \zeta_F$ and $\zeta_n \simeq 0.00015\zeta_F$.
Incidentally, these values almost coincide with simple extrapolations
of the nuclear mass formula to $Z=1,0$: $\zeta_p\simeq - 0.0007$, $\zeta_n
=0$. Since $\zeta_b$ is determined mostly by $\zeta_p$,
$\zeta_m =\zeta_b(\Omega_b/\Omega_m)\sim - 10^{-4}\zeta_F$.
As we have discussed earlier, the constraints from
EDB experiments, as exemplified in Figure \ref{plane}, do not allow
$\alpha$ to change by more than 1 part per billion at red-shifts $z<3.5$.

Restricting our attention to small variations in $\alpha$, we see from
eq. (\ref{al-phi}) that
\be
{\Delta \alpha \over \alpha} = \zeta_F \phi
\label{bm0}
\ee
Then evaluating eq. (\ref{phit}) at $z = 3.5$, we find that
\be
{\Delta \alpha \over \alpha} = {1.2 \over \omega} \zeta_F
\zeta_m \simeq - 10^{-4}\omega^{-1} \zeta_F^2
\label{bm}
\ee
Note that in this model $\zeta_F$ and
$\zeta_m$ are of opposite sign and the final result does not depend on
the sign of $\zeta_F$. Thus from eq. (\ref{bm}), we see that this model
leads to {\em larger} values of
$\alpha$ in the past, which is {\em opposite} the trend reported by Webb
et al.  Moreover, from Figure \ref{plane}, we see that the EDB constraint
requires that $\zeta_F^2/\omega < 10^{-6}$ and thus we see again that
$|\Delta\alpha / \alpha|$ is limited to ${\cal O}(10^{-10})$, in
agreement with the results of \cite{Livio}.
These results however, differ from those of Ref.
\cite{SBM} in both the allowed magnitude and sign of $\Delta\alpha(z<3.5)
/\alpha$.


{\em 2 A String-dilaton-type model}\\

The starting point for this class of models is the action
\be
\int d^4x\sqrt{-g}\exp(-\sqrt{2}\phi)\left(R +  (\partial_\mu\phi)^2 +
\Lambda +{\cal L}_{\rm matter}\right).
\label{strdil}
\ee
The functions $B_i(\phi)$ are easily obtained by making a conformal
transformation to the Einstein frame.  We find that
$\zeta_F = - \sqrt{2}$, $\zeta_\Lambda = \sqrt{2}$, and $\zeta_m =
\sqrt{2}/2$. Furthermore, since there is only one scale in the theory, the
Planck scale, we have $\omega = 1/2$. Therefore, we are able to obtain a
definite value, $\Delta \alpha /\alpha \simeq -3$, (over the redshift
range $z = 3.5 $ to 0.  Clearly this is not realistic and is related to
the well known problem of a massless runaway dilaton in string theory.
Moreover, such a model is ruled out by the EDB constraints, as it predicts
$\zeta_{n,p}\simeq 1$ and $\zeta_n-\zeta_p \sim 10^{-3}$.
Until more can be said about the function $B_\Lambda(\phi)$, there is
no useful way to use string theory to predict changes in $\alpha$.

\newpage
{\em 3 A Brans-Dicke model}\\

In a Brans-Dicke model $\phi$ is initially coupled only to the
gravitational sector of the theory
\be
S_{BD}=\fr{1}{16\pi G_N}\int d^4 x \sqrt{-g}\left[\phi R +\fr{\omega}{\phi}
\partial_\mu \phi \partial^\mu \phi\right ] +S_{\rm matter}
\ee
As in the previous example, it is easy to show
that after the conformal rescaling of metric to the
standard Einstein frame and a field redefinition for $\phi$, the ``new''
field $\phi$ acquires a universal coupling $\zeta$
to the mass sector of the matter fields which is given in terms of
$\omega: \zeta_m = -1/\sqrt{4\omega + 6}$.
One also obtains
$\zeta_\Lambda = - 2\sqrt{2/(2\omega + 3)}$.
As in the Bekenstein model, all physical results depend only on
$\zeta/\sqrt{\omega}$.

Due to conformal invariance of the action for the gauge field,
a tree level coupling of $\phi$ to $F^2$ is absent yielding $\zeta_F = 0$.
However, the conformal symmetry is anomalous, and at the one loop level a
$\phi F^2$ term can be generated. In some sense, the couplings of $\phi$
to the quarks and leptons will be similar (apart from their magnitude)
to those of the Higgs boson. It is then
clear that a non-zero value of $\zeta_F$ will
be generated through the loops of charged particles in the same way that
the Higgs-$\gamma$-$\gamma$ coupling is generated.
For example, the coupling
of the Higgs boson, $h$, to $F^2$ due to the top quark loop can be
obtained by differentiating the top quark contribution to the $QED$
$\beta$-function,
$F^2\ln(\Lambda_{UV}/(m_t(1+h/v)))\rightarrow -F^2 h/v$, where $v$ is the
Higgs vacuum expectation value. This assumes that the ultra-violet cutoff
$\Lambda_{UV}$ is
$h$-independent. In principle, there could be different possibilities
with regard to the $\phi$-dependence of  $\Lambda_{UV}$.
If one assumes that the regulator mass depends on $\phi$ in
exactly the same way as an ordinary mass, then $\phi$ drops out of the loop
amplitude, $\zeta_F$ is not generated and the Brans-Dicke scalar respects
the weak equivalence principle even at the one-loop level.
A different result would arise if we postulate a
$\phi$-independent regularization $\Lambda_{UV}$.
In this case, a non-zero value for $\zeta_F$ is generated and one would
typically have
\be
\zeta_F = -\fr{2\alpha}{\pi}
\zeta_m\left(-\fr{7}{4}+1+2\times \fr{5}{9}\right)
\simeq -1.5 \times 10^{-3} \zeta_m.
\ee
The three terms in parenthesis correspond to the contributions from
$W$-bosons, charged leptons and quarks from second and third generations.
($u$ and $d$ quarks require a separate and quite complicated treatment.
However, their main contribution is given by the charged pion loop and
turns out to be numerically small compared to the contributions of heavy
quarks.) The couplings of $\phi$ to baryons will be simply $\zeta_m$, and
its non-universality appears at the $\zeta_n - \zeta_p \sim
10^{-3}\zeta_m$ level. Thus, the bounds (\ref{EDBl}) push the constraints
on
$\zeta_m^2\omega^{-1}$ to the level of
$10^{-8}$ or so and leave no room for a $O(10^{-5})$ relative
change of $\alpha$ at $0.5\le z \le 3.5$. The maximum allowed change
is not expected to be larger than $10^{-11}$.
The result of Webb et al. cannot be accommodated
in a Brans-Dicke model.

{\em 4 A Supersymmetrized Bekenstein model}\\

While there are many different possible
supersymmetric generalizations of the original Bekenstein model,
we consider the simplest version which begins by promoting
$B_F(\phi)(FF)$ to the rank of a superpotential:
\be
- \int d^4 x {1\over 4} B_F(\phi) F_{\mu\nu}F^{\mu\nu} \rightarrow
\int d^4 x d^2\theta {1\over 4}B_F(\hat\phi) W_\mu W^\mu +({\rm h.c.}).
\label{susyB}.
\ee
Here $\hat \phi$ denotes a chiral superfield, which has $\phi$ as its
bosonic component and $W$ is the supersymmetric field strength.
In component notation this interaction can
be rewritten as
\be
\int d^4 x \left(B_F(\phi)\left[-{1\over 4}F_{\mu\nu}F_{\mu\nu}+
{1\over 2}\bar\chi \partial_\mu \bar\gamma_\mu \chi \right ]
-\fr{1}{2}B^\prime_F(\phi) F_\phi \chi^T\chi \right)...
\label{comp}
\ee
$F_\phi$ denotes the $F$-component of $\hat\phi$ and
ellipses stand for other terms not relevant for the present discussion.
We see that in addition to the interaction with the gauge boson,
$\phi$ acquires a coupling to the gauge fermion or gaugino, $\chi$.
$F_\phi$ may acquire a v.e.v. which contributes to the
supersymmetry-breaking gaugino mass. There may also be
additional soft-breaking contributions leading to a mass term of the form
${1\over 2}M \chi^T\chi$. Performing the rescaling $\chi \rightarrow
\chi/\sqrt{B_F(\phi)}$, we arrive at the following Lagrangian in the
$\phi-\chi$ sector,
\be
{\cal L}_{\phi\chi} = {1\over 2}\bar\chi \partial_\mu \bar\gamma_\mu \chi
-\fr{1}{2}\fr{B^\prime_F(\phi) \langle F_\phi\rangle + M}{B_F(\phi)}
\chi^T\chi
\ee
In the linearized version of the theory given by eqs. (\ref{expansion}),
we arrive at the following expression for $\zeta_\chi$,
\be
\zeta_\chi = \fr{(\xi_F-\zeta^2_F) \langle F_\phi\rangle -  M\zeta_F }{M_\chi},
\label{zetaSB}
\ee
where $M_\chi = M +\zeta_F\langle F_\phi\rangle$.

Clearly, $\zeta_\chi$ can be $O(1)$, if $\zeta_F\sim O(1)$, however its
sign is not uniquely defined unless we make some specific
assumptions about $\zeta_F$, $\xi_F$, $\langle F_\phi\rangle$ and $M$.
For example, let us take $B_F$ as in the original Bekenstein model so
that
$\zeta_F = -2$.  Let us further assume that supersymmetry breaking occurs
outside the $\phi$ sector so that $F_\phi = 0$.  In this case, $M_\chi =
M$ and $\zeta_\chi = 2$.  Since the dark matter dominates the energy
density of non-relativistic matter, we have
$\zeta_m \simeq  \zeta_\chi$. Indeed, it is quite reasonable to
expect that in general $|\zeta_m|\sim |\zeta_F|$ which leads to the
relation between the dark matter and baryonic sources of $\phi$ advertized
in  (\ref{intro}).

The final parameter which must be specified in the model is $\omega$.
In order to obtain consistency with the
combination of EDB constraints, we must have $|\zeta_F/\sqrt{\omega}|<
10^{-3}$ or $\omega > 4 \times 10^6$. If we again assume $\zeta_\Lambda =
0$, we can compute the change in the fine structure constant (from $z =
3.5$ to $z=0$)
\be
{\Delta \alpha \over \alpha} = {1.2 \over \omega} \zeta_F
\zeta_m \simeq -5/\omega
\ee
Note again the sign of $\Delta \alpha$ predicts that $\alpha$ was larger
in the past, although this conclusion could be modified if $\phi$
contributes to supersymmetry breaking (so that $F_\phi \ne 0$).
 Also, because of the EDB constraint, the relative change of $|\Delta
\alpha/\alpha|$ at $z$ in the interval $0.5-3.5$ would typically be at
the level of $10^{-6}$, unless some additional fine-tuning is
introduced. (For example, if a partial cancellation between $M$ and
$\zeta_F\langle F_\phi\rangle$ in $M_\chi$ occurs, one can get
$|\zeta_m|>|\zeta_F|$ and thus satisfy the constraints shown in Figure 2.)

In contrast with the non-suspersymmetric version of the Bekenstein model,
the change in $\phi$ from the time of the radiation domination/matter
domination transition to the present epoch can be of order 1 or even
larger. In this case, obviously, the linearized approach to
$B_i(\phi)$ may fail for $z \gg 1$. Therefore, it is impossible
to determine the total change of $\alpha$ from the BBN epoch, without
specifying the complete functional form for both $B_\chi(\phi)$ and
$B_F(\phi)$. Nevertheless, it can be
shown that if the $B_i(\phi)$ are dominated by the few first terms in the
Taylor expansion up to $z\sim 10^5$, the change of $\alpha$ is within
the BBN bounds. Large changes in $\phi$ may also entail
a non-negligible backreaction
of the $\phi$-dependent stress-energy tensor on Freidman's equations. In
this case, one could get interesting effects in the expansion of the
Universe due to the $B_\chi(\phi)M_\chi \bar \chi \chi$ term, which
can be interpreted at the same time as varying mass dark matter
\cite{AC} or the potential term for $\phi$, that has an overall factor of
$\rho_m$.

\begin{table}
\begin{center}
\begin{tabular}{|c|c|c|c|c|c|}\hline & &&&&\\
    Type of model & $\zeta_F$ & $\zeta_b$ & $\zeta_m$
&$\fr{\Delta g}{\bar g}$ & $|\fr{\Delta\alpha}{\alpha}|_{\rm max}$ \\
 & &&&& at $0.5\le z\le 3.5$\\\hline \hline & &&&& \\
 Bekenstein model &$-2$ &$10^{-3}$ &$10^{-4}$ &
$\fr{10^{-6}}{\omega}$&$10^{-10}$
  \\   & &&&&  \\
 ``String dilaton''$^*$ & $-\sqrt{2}$ &$-\sqrt{2}/2$&$-\sqrt{2}/2$&
$10^{-3}$ & $ 1$ \\ &
&&&&  \\
Brans-Dicke model$^{**}$   &0 &$-1/\sqrt{4\omega + 6}$&$-1/\sqrt{4\omega
+ 6}$&$\fr{10^{-3}}{\omega^2}$& $10^{-11}$ \\ & &&&&
\\ SUSY BM   &$ 1$ & $10^{-3}$ &$
1$&$\fr{10^{-6}}{\omega}$
 & $10^{-6}$ \\ &&&&&  \\
$M_\chi$-driven   & $10^{-4}$ & $ 10^{-7}- 10^{-5}$ & $ 0.1$  &
$\fr{10^{-12}}{\omega}$ & $10^{-5}-10^{-4}$
\\& &&&& \\\hline
\end{tabular}
\caption{Order of magnitude model predictions for the set of relevant
couplings, non-universality of $\phi$-exchange, and maximum allowed
$|\Delta\alpha/\alpha|$ at
$0.5\le z\le 3.5$.
(*) The tree-level form of $B_i(\phi)$ is assumed. (**)
A $\phi$-dependent cutoff is assumed.}
\end{center}
\end{table}

An interesting consequence of models where $\phi$ couples to dark matter
is the non-universality of the free fall towards an attractor dominated by
dark matter, e.g. the center of a galactic halo. The magnitude of
differential acceleration of a system of heavy/light
elements towards a dark matter
attractor is enhanced compared to the acceleration towards the Sun
by the factor $\zeta_m/\zeta_p$:
\be
 \fr{\Delta g}{\bar g}=\fr{\zeta_m}{\omega}
\left[(\zeta_n-\zeta_p)\left(\fr{Z_1}{A_1}-\fr{Z_2}{A_2}\right)+
7 \times 10^{-4}\zeta_F\left(\fr{Z_2^2}{A_2^{4/3}}
-\fr{Z_1^2}{A_2^{4/3}}\right)\right] \simeq
3\times 10^{-2} \fr{\zeta_m\zeta_F}{\omega}
\ee
Depending on the ratio of $\zeta_m/\zeta_p$, this effect may be as
large as $10^{-7}-10^{-9}$ when the usual EDB constraints
are satisfied. Unfortunately, this is much lower than the
current experimental sensitivity achieved, $\sim 0(10^{-3})$, in 
tests of differential acceleration towards galactic center \cite{AdelDM}.

\vskip .3in 

{\em 5 A gaugino driven modulus }

It may happen that the modulus field that changes $\alpha$ is coupled
primarily to the soft breaking parameters. Then the coupling to $F^2$ and
baryons may only appear at the loop level. Let us suppose for simplicity that
initially $\phi$ couples only to gaugino masses,
\be
{\cal L}= \sum_{i=1,2,3}\left[\fr{1}{2}\bar \lambda_i \partial_\mu \gamma^\mu
\lambda_i - \fr{1}{2} M_i(1+\zeta_{M_i}\phi)\lambda^T\lambda\right],
\ee
where the summation is over the three Standard Model gauge groups
(color and weak indices are suppressed).
We assume that the lightest supersymmetric particle is the neutralino $\chi$
which is predominantly the bino. Therefore, $M_\chi\simeq M_1$ and
$\zeta_m\simeq \zeta_\chi \simeq \zeta_{M_1}$.

We now consider the possibility that all couplings of $\phi$ to
standard model fields are induced radiatively. At the one loop
level, the couplings with $SU(2)$ and $SU(3)$ gauge bosons will be
generated,
\ba
\zeta_W = -\fr{2\alpha_W}{3\pi} \zeta_{M_2}
\\
\zeta_G = -\fr{\alpha_s}{\pi} \zeta_{M_3}
\ea
The calculation of these couplings is trivial: they are obtained  by
differentiating gluino and wino contributions to the corresponding
beta functions over $M_i$. In the derivation of these couplings we
assumed that the cutoff scale is $\phi$-independent.

After the breaking of the $SU(2)\times U(1)$ gauge symmetry, $\zeta_W$
induces a contribution to $\zeta_F$,
\be
\zeta_F = \sin^2\theta_W\zeta_W \simeq - 1.5 \times 10^{-3} \zeta_{M_2}.
\ee
The coupling to baryons is mediated by $\zeta_F$ as before or
by $\zeta_G$ or by the $\phi m_q \bar qq $ operators, induced at the
supersymmetric threshold. Typically,
$\zeta_G$ induces too large a coupling to baryons, $\zeta_{n,p} \sim
(0.02-0.06)\zeta_{M_3}$, to be consistent with EDB limits and $\omega =
O(1)$. Therefore, one must require  $\zeta_{M_3} \ll
\zeta_{M_1},\zeta_{M_2}$. Wilson coefficients in front of
supersymmetric threshold-induced  $\phi m_q \bar qq $ operators
are expected to be at the level of $10^{-5}-10^{-3}$ from
$\zeta_{M_1}$ and $\zeta_{M_2}$. This creates a coupling of $\phi$
to nucleons at the level $\zeta_{n,p}\sim (10^{-6}-10^{-4})\zeta_{M_2}$,
which is mostly due to a large matrix element of the $m_s\bar ss$ operator
and/or the two-loop induced $G^a_{\mu\nu}G^{a\mu\nu}$ operators.
$m_u\bar uu$ and $m_d \bar dd$
generate the difference between the couplings to
neutron and proton at the level
of $\zeta_{n}-\zeta_{p}\sim (10^{-7}-10^{-5})\zeta_{M_2}$. Due to the
small values of $\zeta_{p,n}, \zeta_p - \zeta_n$, and $\zeta_F$,   the
constraints based on the violation of the equivalence principle
(\ref{EDBl}) do not lead to very restrictive bounds,
\be
\zeta_{M_2}^2 / \omega \la 10^{-4}-10^{-2}.
\label{mchilim}
\ee
The possibility of ``choosing'' $\zeta_{M_1}$, $\zeta_{M_2}$
and $\omega$ creates sufficient freedom to satisfy EDB constraints and
at the same time have $\Delta\alpha/\alpha$ at $0.5\le z \le 3.5$
compatible with the Webb et al. result.
Recall that
\be
{\Delta \alpha \over \alpha} = {1.2 \over \omega} \zeta_F
\zeta_m \simeq - {1.8 \times 10^{-3} \over \omega} \zeta_{M_1}
\zeta_{M_2}
\ee
Indeed, to satisfy both it is
sufficient to take
\be
\fr{\zeta_{M1}}{\sqrt{\omega}} = -  \fr{\zeta_{M2}}{\sqrt{\omega}} \simeq \pm
0.1;\;\;\;\;\;\; {\rm and}\;\;\;\;\;\; \zeta_{M3} \ll \zeta_{M1},\zeta_{M2}.
\ee

To conclude this section, we combine all model predictions in Table 1.

\section{Models of an oscillating fine structure constant}
\setcounter{equation}{0}

Finally,  we turn to the case when all of the
functions $B_i(\phi)$ have a common extremum point $\phi_{\rm ext}$.
As was shown by Damour and Nordtvedt and Damour and Polyakov \cite{DP},
the matter energy density may serve as a cosmological attractor for $\phi$,
so that today  its value is close to $\phi_{\rm ext}$. In our approach,
without loosing generality, we choose $\phi_{\rm ext}=0$. The requirement
of a common extremum is equivalent to condition that ${\em all}$ linear
couplings $\zeta_i = 0$.

The cosmological evolution of $\phi$ is now given by the
$\xi_i$ couplings. There are two distinct regimes to consider:
$\phi(t)=const$ at early times and an oscillating or runaway regime at
late times. These two regimes are common for cosmological evolution of any
quasi-modulus field, e.g. axion. The transition occurs when the Hubble
rate drops below the effective (time-dependent) mass of $\phi$,
\be
m_\phi^2 =
{2\over\omega}\left[\Lambda_0\xi_\Lambda + \fr{\rho_m\xi_m}{M_{\rm Pl}^2}
\right]
={6 H_0^2\over\omega}\left[\Omega_\Lambda\xi_\Lambda +\Omega_m\xi_m
\left ({a_0\over a }\right)^3
\right ]
\label{massofphi}
\ee
The sign of $m_\phi^2$ determines if it is a runaway or oscillatory
evolution. Here, we are interested in the
oscillatory regime, and thus assume that  $m_\phi^2$ is positive. The
amplitude of these oscillations red-shifts as $a^{-\kappa}$, where
$3/4<\kappa<3/2$. $\kappa = 3/2$ occurs if (\ref{massofphi}) is
dominated by the first term, i.e. rigid mass, and $\kappa = 3/4$
if the second matter-induced term is dominant \cite{DP}. Thus,
the effective value of $\alpha$ also oscillates, at twice the frequency,
$2m_\phi$, and with an amplitude decreasing as $a^{-2\kappa}$.
In this regime, is it possible to satisfy the EDB and Oklo bounds and have
$\Delta \alpha/\alpha \sim 10^{-5}$ at $ 0.5 \le z\le 3.5$?


If $\xi_F$ is the dominant source of
the couplings to baryons, the expected level of the violation of
the equivalence principle is
\be
\fr{\Delta g}{\bar g} \simeq 10^{-6} \fr{\xi_F^2\phi_{\rm now}^2}
{ \omega}
\label{dg}
\ee
Of course, it is possible that the value of $\phi$ today is close to
zero, simply because in the oscillatory regime, $\phi=0$ occurs
regularly. This is, however, an accidental situation, and one would
naturally expect
$\phi_{\rm now}$ to be on the order of the amplitude of oscillations.
On the other hand, the relative change of $\alpha$ is given by
\be
\fr{\Delta \alpha}{\alpha} = \fr{1}{2}\xi_F(\phi^2(z) - \phi^2_{\rm now})
\simeq  \fr{1}{2}\xi_F\phi^2(z)
\ee
Using the relation between $\phi(z)$ and $\phi_{\rm now}$, and
plugging in the constraint from (\ref{dg}), we get
\be
\fr{\Delta\alpha}{\alpha} \la 10^{-5} (1+z)^{2\kappa}{\omega\over \xi_F}
\ee
It is then clear that this can be consistent with $10^{-5}$ at
$0.5\le z\le 3.5$ naturally without a fine-tuning of parameters when
$\omega/\xi_F \sim 1$.
The Oklo bounds can be made marginally
consistent with \cite{Webb01} in this scenario
only for large $z$ (close to 3.5 rather than 0.5) and for large $\kappa$,
$\kappa= 3/2$. This favors models where the oscillations of $\phi$
at $z <3.5$ are driven by the rigid $\xi_\Lambda$-proportional mass term
of $\phi$.

\section{Conclusions}

We have shown that generalized versions of the Bekenstein's model may be
consistent with the strong limits imposed by E\"otv\"os-Dicke-Braginsky
type of experiments and at the same time provide a relative change of
$\alpha$ at the $10^{-5}$-level, claimed recently in Webb et al.
\cite{Webb01}. The necessary flexibility in our models is achieved by
the coupling of the modulus field $\phi$ to the dark matter energy density
and to the cosmological constant. We argue that it is natural to expect
that the cosmological evolution of $\phi$ will be mostly driven by these
sources rather than by the baryon energy density. This can be
seen explicitly in the simplest SUSY-version of the Bekenstein model,
where the supersymmetric partner of the $U(1)$ gauge field is the
dominant non-baryonic component of dark matter.

In practice, it turns out that among various models where $\phi$ couples to
$F^2$, baryons and dark matter, only a few survive the EDB constraints and
provide the $O(10^{-5})$ relative change in $\alpha$ over the redshift
range
$0.5\le z\le 3.5$. In particular, we find that the models where $\phi$ is
coupled initially only to $U(1)$ and $SU(2)$ gaugino mass terms can easily
satisfy both criteria.

The bounds on $\Delta\alpha$ from the Oklo phenomenon are less dependent
on the details of the coupling of $\phi$ to the matter field. Generally,
they are strong enough to rule out the change of the fine structure
constant, implied by Webb et al. In the context of the generalized models
discussed here, the negative coupling of $\phi$
to the cosmological constant may be used to slow down its evolution and
make Oklo bounds consistent with \cite{Webb01}. This possibility, however,
looks accidental and very fine-tuned for  $\Delta\alpha/\alpha \sim 10^{-5}$
at $ z= 0.5$.
Of course, our treatment of all models at the loop level is
plagued by the usual problem of the cosmological constant and
the near-masslessness of the moduli field $\phi$. This prevents
us from making any prediction for the size of the $\zeta_\Lambda$
coupling constant. We also find that $\zeta_i=0$ \cite{DP} is easier to
reconcile with EDB constraints and Webb et al., as in this case there is
an additional suppression of the $\phi$-mediated force.

\noindent{ {\bf Acknowledgments} } \\
\noindent  
This work was supported in part by DOE grant
DE--FG02--94ER--40823 at the University of Minnesota
and NSERC.
M.P. would like to thank the theory
groups at McGill University and UQAM for their warm hospitality.
We benefited from conversations with R. Cyburt, D. Demir,
B. Fields, T. ter Veldhuis and especially M. Voloshin.



\begin{thebibliography}{99}

\bibitem{Dirac}
P.A.M. Dirac, Nature, {\bf 139} (1937) 323.

\bibitem{Sister}
P.~Sisterna and H.~Vucetich,
Phys.\ Rev.\ D {\bf 41} (1990) 1034.

\bibitem{Bek}
J.~D.~Bekenstein,
Phys.\ Rev.\ D {\bf 25} (1982) 1527.

\bibitem{EDB} R.V. E\"otv\"os, V. Pekar and E. Fekete, Ann. Phys. (Leipzig)
{\bf 68} (1922) 11; \\P.~G.~Roll, R.~Krotkov and R.~H.~Dicke,
Annals Phys.\  {\bf 26} (1964) 442;\\V.B. Braginsky and V.I. Panov,
Zh. Eksp. Teor. Fiz. {\bf 61} (1972) 873 [Sov. Phys. JETP {\bf 34}
(1972) 463.

\bi{Livio} M. Livio and M. Stiavelli, Ap. J. Lett. {\bf 507} (1998) L13.

\bi{Webb01} J.K. Webb {\em et al.},  Phys. Rev. Lett. {\bf 87} (2001) 091301.

\bi{Oklo} A. I. Shlyakhter, Nature {\bf 264} (1976) 340.

\bi{DD} T.~Damour and F.~Dyson,
Nucl.\ Phys.\ B {\bf 480} (1996) 37.

\bi{Fujii} Y. Fujii {\em et al.}, Nucl. Phys. {\bf B573} (2000) 377.

\bi{LV} S.~J.~Landau and H.~Vucetich,
astro-ph/0005316; N.~Chamoun, S.~J.~Landau and H.~Vucetich,
Phys.\ Lett.\ B {\bf 504} (2001) 1.

\bibitem{bbn} E.~W.~Kolb, M.~J.~Perry and T.~P.~Walker,
Phys.\ Rev.\ D {\bf 33}, 869 (1986);
B.~A.~Campbell and K.~A.~Olive,
Phys.\ Lett.\ B {\bf 345}, 429 (1995)
[arXiv:hep-ph/9411272];
L.~Bergstrom, S.~Iguri and H.~Rubinstein,
Phys.\ Rev.\ D {\bf 60}, 045005 (1999)
[arXiv:astro-ph/9902157].


\bibitem{SBM}
H.~B.~Sandvik, J.~D.~Barrow and J.~Magueijo,
arXiv:astro-ph/0107512;
J.~D.~Barrow, H.~B.~Sandvik and J.~Magueijo,
arXiv:astro-ph/0109414.



\bi{Sacha} S. Davidson, B. Campbell and D. Bailey, Phys. Rev.{\bf D43} (1991)
2314.

\bi{PtV} M.~Pospelov and T.~ter Veldhuis,
Phys.\ Lett.\ B {\bf 480} (2000) 181
[hep-ph/0003010].

\bi{DP} T.~Damour and K.~Nordtvedt,
Phys.\ Rev.\ Lett.\  {\bf 70} (1993) 2217;\\
T.~Damour and K.~Nordtvedt,
Phys.\ Rev.\ D {\bf 48} (1993) 3436;\\
T.~Damour and A.~M.~Polyakov,
Nucl.\ Phys.\ B {\bf 423} (1994) 532
[hep-th/9401069].

\bi{nko}N.~Kaloper and K.~A.~Olive,
Astropart.\ Phys.\  {\bf 1} (1993) 185.

\bibitem{tsey} A.A. Tseytlin,
Mod. Phys. Lett. {\bf A6} (1991) 1721;
Class. Quant. Grav. {\bf 9} (1992) 979;
Int. J. Mod. Phys. {\bf D1} (1992) 223.

\bi{Dicke} R. H. Dicke, {\em The Theoretical Significance of Experimental
Relativity} (Gordon and Breach, New York, 1965).

\bi{LunarR} J.O. Dickey {\em et al.}, Science {\bf 265} (1994) 482;\\
J.G. Williams {\em et al.}, Phys. Rev. {\bf D53} (1996) 6730;\\
T.~Damour, arXiv:gr-qc/0109063.

\bi{DZ}G.~R.~Dvali and M.~Zaldarriaga,
hep-ph/0108217.
For some unknown reasons, the (dominant) contribution from the
nuclear electromagnetic energy was not taken into account.

\bi{W}S.~Weinberg,
Rev.\ Mod.\ Phys.\  {\bf 61} (1989) 1.


\bi{GL}
J.~Gasser and H.~Leutwyler,
Phys.\ Rept.\  {\bf 87} (1982) 77.

\bibitem{AC} G.~W.~Anderson and S.~M.~Carroll,
arXiv:astro-ph/9711288.

\bi{AdelDM} G. Smith {\em et al.}, Phys. Rev. Lett. {\bf 70} (1993) 123.


\end{thebibliography}
\end{document}